\documentstyle[seceq,preprint]{jpsj}

\def\runtitle{Corner Transfer Matrix Algorithm for Classical
Renormalization Groups}
\def\runauthor{Tomotoshi {\sc Nishino} and Kouichi {\sc Okunishi}}

\title{ Corner Transfer Matrix Algorithm for
Classical Renormalization Group}

\author
{ Tomotoshi {\sc Nishino}\footnote{e-mail:
nishino@phys.kobe-u.ac.jp} and
Kouichi {\sc Okunishi}$^{1,}$\footnote{e-mail:
okunishi@godzilla.phys.sci.osaka-u.ac.jp} }

\inst
{ Department of Physics , Faculty of Science,
Kobe University, Rokkodai 657\\
$^1$Department of Physics, Graduate School of Science,
Osaka University, Toyonaka 560 }

\abst{We report a real-space renormalization group (RSRG) algorithm,
which is formulated through Baxter's corner transfer matrix (CTM), for
two-dimensional ($d = 2$) classical lattice models. The new method
performs the renormalization group transformation according to White's
density matrix algorithm, so that variational free energies are
minimized within a restricted degree of freedom $m$. As a consequence
of the renormalization, spin variables on each corner of CTM are
replaced by a $m$-state block spin variable. It is shown that the
thermodynamic functions and critical exponents of the $q = 2, 3$ Potts
models can be precisely evaluated by use of the renormalization group
method.}

\kword
{ Density Matrix, Corner Transfer Matrix, Renormalization Group. }

\begin{document}
\sloppy
\maketitle

\section{Introduction}

The density matrix renormalization group (DMRG) established by
White~\cite{Wh1,Wh2} has greatly enhanced the applicability of the
numerical real-space renormalization group (RSRG)~\cite{Kad,Wi,Rs} to
one-dimensional ($d = 1$) quantum systems. After the applications to
the $S = 1$ Heisenberg spin chain,~\cite{Wh2,Huse,Sol} DMRG has been
applied to a number of $d = 1$ quantum systems, such as Heisenberg
ladder,~\cite{Hida,Wh25,Sch,Wh3} bond-alternating spin
chain,~\cite{Kato,Yajima} strongly correlated electron
system,~\cite{Yu,Noa,Shibata,Wh4} etc. On one hand improvements upon
the numerical algorithm of DMRG have been proposed in order to analyze
impurity systems,~\cite{Saf,WangX} random systems,~\cite{Hida2,HidaN}
Bethe lattice systems,~\cite{Otsuka} correlated electron system
defined in the momentum space,~\cite{Xiang} spin chains under finite
temperature,~\cite{Xiangf} and so on.

Recently White has analyzed the ground state wave function of $d = 2$
quantum spin systems, using a numerically accelerated finite system
DMRG algorithm.~\cite{Wh4} At first, White's acceleration technique
was only applicable to the finite system DMRG algorithm. Soon after
the authors extended the acceleration technique to infinite system
DMRG algorithm.~\cite{PWF} Further numerical acceleration in DMRG has
been reported for models that have (quantum-) group
symmetry.~\cite{IRF}

The DMRG picks up relevant correlations between the local system
($=$block) and the rest of the system ($=$reservoir). Since irrelevant spin (or
particle) fluctuations are projected out, the DMRG method
intrinsically has a variational property. Indeed, the ground state
wave function obtained by DMRG is a good variational wave function,
which is written as a product of tensors.~\cite{Zi,Zi2,Zi3} In the
thermodynamic limit the tensors become position independent, where the
DMRG coincides with \"Ostlund's variational
method.~\cite{Ostlund,Ostlund2} Mart\`{\i}n-Delgado, Rodriguez-Laguna
and Sierra have reformulated the variational relation in DMRG using
projection operators, and have proposed new RSRG
algorithms.~\cite{Si1,Si2,Si3}

Since $d = 1$ quantum systems are naturally related to $d = 2$
classical systems, it is possible to apply DMRG algorithm to the
latter.~\cite{Ni} The largest eigenvalue of the row-to-row transfer
matrix is primarily important for the analysis of the $d = 2$
classical models. The DMRG applied to a $d = 2$ classical model
evaluates the lower bound of the largest eigenvalue,  using a
variational state vector written in a product of tensors. It has been
shown that the thermodynamic functions can be obtained precisely by
DMRG in off critical regions. However, the numerical convergence in
free energy becomes slow near the critical temperature, and therefore
extensive computations are necessary at criticality. The reason of
this slowing down is that the maximum eigenvalue of the row-to-row
transfer matrix is nearly degenerate in the critical region. Such a
degeneracy spoils the numerical efficiency of the Lanczos
diagonalization, that is the most time consuming part in DMRG.

Baxter's method of corner transfer matrix~\cite{Bx1,Bx2,Bx3} (CTM),
that was formulated in 1968 as an extension of the Kramers-Wannier
approximation,~\cite{Krm,Kik} is another variational method for $d =
2$ classical lattice systems. The method gives approximate free energy
per site in the thermodynamic limit. Baxter's method seems to
have no relation with DMRG, but actually both of them are deeply
connected; they are both iterative renormalization group method, and
have the same fixed point in the thermodynamic limit. Baxter's method
can be used as a numerical method,~\cite{Bx1,Bx2,Bx3} and it runs
faster than DMRG at criticality. This is because the largest
eigenvalue of CTM is not degenerate even at the critical point.

In this paper we introduce the advantage of Baxter's method into the
numerical algorithm of DMRG for $d = 2$ classical system. We express
the density matrix as a product of four CTMs. For the brevity, we call
the improved renormalization group method `corner transfer matrix
renormalization group' (CTMRG) in the following.~\cite{CTMRG} Apart
from Baxter's method, the purpose of CTMRG is to obtain variational
free energies of {\it finite size systems}, up to a certain system
size.

In the next section we review the construction of CTM. In order to
simplify the discussion, we consider a $q$-state Potts model on a
decorated lattice, because it is easy to define CTM on the lattice. In
\S 3 we explain the variational relation in DMRG using projection
operators. We then introduce CTM into the formulation of DMRG, and
make up the numerical algorithm of CTMRG in \S 4. In \S 5 we apply
CTMRG to $q = 2, 3$ Potts models. It is verified via the finite size
scaling analysis~\cite{Fi,Ba} that CTMRG gives correct critical
exponents. Conclusions are summarized in \S 6.

\section{Corner Transfer Matrix}

We consider a $q$-state Potts model~\cite{Potts,WuP} on a decorated
square lattice, whose geometry is shown in Fig.1. The white marks
represent $q$-state spins $\{s\}$ on vertices, and the black ones
represent another set of $q$-state spins $\{\sigma\}$ that are in
between $s$-spins. We refer to the model as `decorated Potts model' in
the following.

\begin{figure}
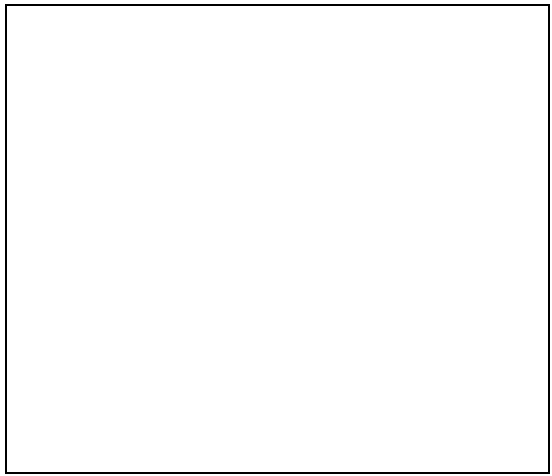

\figureheight{6cm}
\caption{The $q$-state Potts model on a decorated lattice. The black
and white marks represent $q$-state spin variables $\{\sigma\}$ and
$\{s\}$, respectively.}
\label{fig:1}
\end{figure}

We consider a square shaped finite size system of linear dimension
$L$; the case $L = 3$ is shown in Fig.1. The partition function is
\begin{equation}
Z^L = \sum_{\{\sigma\}} \sum_{\{s\}} {\rm exp}
\bigg\{ K^* \sum_{\langle ij\rangle} \delta(\sigma_i,s_j) \bigg\},
\end{equation}
where $K^*$ is the interaction parameter, $\langle ij\rangle$
specifies the neighboring $\sigma$-$s$ spin pairs, and
$\delta(\sigma_i,s_j)$ is equal to unity if $\sigma_i = s_j$ and zero
otherwise. When $q = 2$ the model coincides with the super exchange
Ising model by Fisher.~\cite{Fis} Since the decorated lattice is
bipartite, we can explicitly take the configuration sum over
$\sigma$-spins, leaving that for $s$-spins. The partition function
after the summation is
\begin{equation}
Z^L = \big( q - 2 + 2e^{K^*} \big)^M \sum_{\{s\}}  {\rm exp}
\bigg\{ K \sum_{\langle ij\rangle} \delta(s_i,s_j) \bigg\},
\end{equation}
where $M = 2L(L-1)$ is the number of $\sigma$-spins in the system of
the linear dimension $L$, and $K$ is the effective interaction
parameter between $s$-spins; $K$ is determined through the duality
relation for the $d = 1$ Potts model~\cite{WuP}
\begin{equation}
e^K = { {q - 1 + e^{2K^*}} \over {q - 2 + 2e^{K^*}} } .
\end{equation}
Equation (2.2) shows that $Z^L$ is proportional to the partition
function of the Potts model on the simple square lattice.

In the same manner, we can decimate $s$-spins and express $Z^L$ in
eq.(2.1) as a partition function of a symmetric vertex model. The
vertex weight, that represents four $\sigma$-spin interaction around a
vertex, is expressed as
\begin{equation}
W_{abcd}^{~} = \sum_{s=1}^q {\rm exp}
\bigg\{ K^*( \delta_{sa} + \delta_{sb}
+ \delta_{sc} + \delta_{sd} ) \bigg\},
\end{equation}
where $s$ represents an $s$-spin on the vertex, and $a, b, c, d$
represent the $\sigma$-spins around $s$. The weight $W_{abcd}$ is
invariant under the permutations of indices, since our model is
isotropic. (See eq.(2.1).) We refer to the vertex model as
`$q^4$-vertex model', since the indices $a, b, c,$ and $d$ run from
$1$ to $q$, and since all of the weights are nonzero.

In order to express $Z^L$ correctly, we have to define two additional
Boltzmann weights at the boundary. One is
\begin{equation}
P_{abc}^{~} = \sum_{s=1}^q {\rm exp}
\bigg\{ K^*( \delta_{sa} + \delta_{sb} + \delta_{sc}) \bigg\}
\end{equation}
that expresses the three-spin interaction at the side of the square
system, and the other is
\begin{equation}
C_{ab}^{~} = \sum_{s=1}^q {\rm exp}
\bigg\{ K^*( \delta_{sa} + \delta_{sb}) \bigg\}
\end{equation}
that represents two-spin interaction at the corner. The partition
function $Z^L$ in eq.(2.1) is expressed as a tensor product of these
Boltzmann weights; for example, the partition function when $L = 3$
is
\begin{equation}
Z^3 = \sum_{ab\ldots l} W_{kheb}^{~} P_{abc}^{~}
C_{cd}^{~} P_{def}^{~}
C_{fg}^{~} P_{ghi}^{~} C_{ij}^{~} P_{jkl}^{~} C_{la}^{~},
\end{equation}
where the arrangement of spin indices is shown in Fig.2.

\begin{figure}
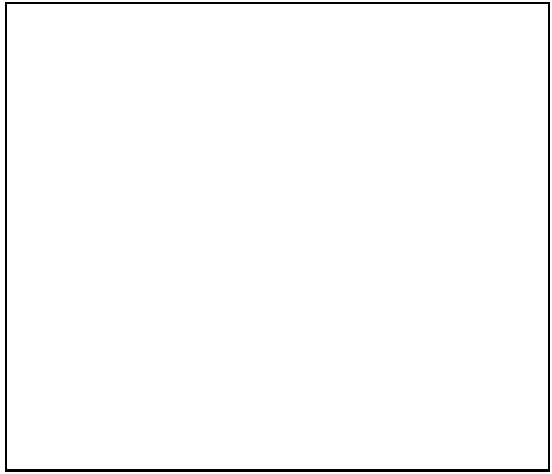

\figureheight{6cm}
\caption{The symmetric $q^4$-vertex model that corresponds to the
decorated $q$-state Potts model in Fig.1.}
\label{fig:2}
\end{figure}

In order to generalize the expression of $Z^3$ in eq.(2.7) to larger
systems, we introduce the half-row transfer matrix~\cite{Hr} (HRTM)
and CTM. The HRTM is the left (or the right) half of the row-to-row
transfer matrix, which is a generalization of the boundary weight in
eq.(2.5). The HRTM of length $N$ can be defined by the recursion
relation
\begin{equation}
P^N_{{\mib a}b{\mib c}}
= \sum_{d=1}^q W_{a_N\,\,d\,\,c_N\,\,b}^{~} \,
P^{N-1}_{{\mib a'}d\,{\mib c'}} ,
\end{equation}
where the vector index
\begin{equation}
{\mib a} = (a_1, a_2,\ldots, a_{N-1}, a_N)
\end{equation}
represents a group of $q$-state spin indices on the half-row of
length $N$, and ${\mib a}' =  (a_1, \ldots, a_{N-1})$ is included in
${\mib a} = ({\mib a}', a_N)$; the same for ${\mib c} = ({\mib c'},
c_N) $. The initial value $P^{\,1}_{{\mib a}b{\mib c}}$, where ${\mib
a} = (a_1)$ and ${\mib c} = (c_1)$, is given by the boundary weight in
eq.(2.5). As an example, we show the case $N = 3$ in Fig.3. We often
abbreviate the vector indices of $P^N_{{\mib a}b{\mib c}}$ and write
the HRTM as $P^N_b$; in such a case we think of $P^N_b$ as a
$q^N$-dimensional matrix $(P^N_b)_{{\mib a}{\mib c}}$.

\begin{figure}
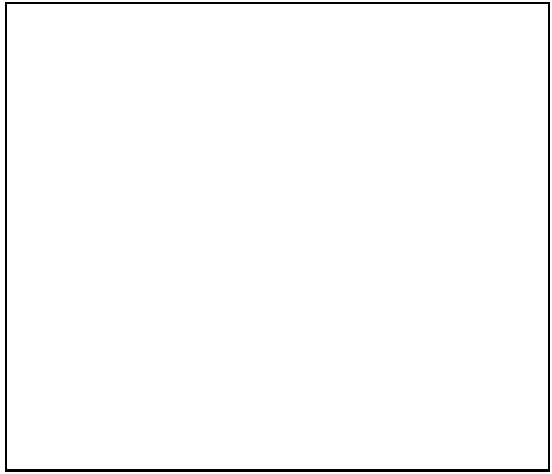

\figureheight{6cm}
\caption{The half-row transfer matrix of length 3 ($= P^3$), and its
components $P^2$ and $W$.}
\label{fig:3}
\end{figure}

The CTM is a generalization of the boundary weight $C_{ab}$ in
eq.(2.6). We define CTM using the recursion relation
\begin{equation}
C^N_{{\mib a}{\mib b}} = \sum_{{\mib c'}{\mib d'}} \bigg(
\sum_{ef} W_{e\,\,f\,\,b_N\,\,a_N}^{~} \,
P^{N-1}_{{\mib a'}e\,{\mib c'}} \,
P^{N-1}_{{\mib b'}f\,{\mib d'}} \,  \bigg)
C^{N-1}_{{\mib c'}{\mib d'}} \, ,
\end{equation}
where we have used the index rule ${\mib a} = ({\mib a'}, a_N)$ in
eq.(2.9); the same for ${\mib b}, {\mib c}$ and ${\mib d}$. The
initial value $C^{\,1}_{{\mib a}{\mib b}}$, where ${\mib a} = (a_1)$
and ${\mib b} = (b_1)$, is given by the corner weight in eq.(2.6).
Figure 4 shows the CTM when $N = 3$. The factor $W P_e^{N-1}
P_f^{N-1}$ inside the parenthesis in eq.(2.10) plays the role of a
transfer matrix that extends the area of CTM.

It should be noted that our definition of CTM in eq.(2.10) is
different from the conventional one; $C^N$ in eq.(2.10) corresponds to
a square quadrant of the whole system, while Baxter's CTM corresponds
to a triangular region of a square lattice system.~\cite{Bx3} Because
of this difference, two HRTMs are necessary when we increase the size
of CTM. (See eq.(2.10))

\begin{figure}
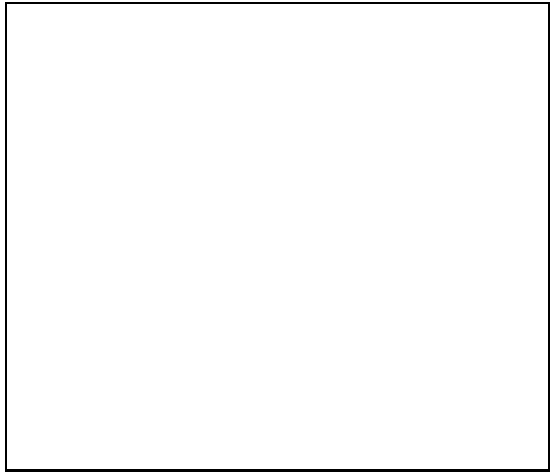

\figureheight{6cm}
\caption{Corner Transfer Matrix $C^3$, and its components $P^2$,
$C^2$, and $W$. The shaded region corresponds to $C^2$.}
\label{fig:4}
\end{figure}

We have so far chosen free boundary conditions. Since the boundary
weights  $P_{abc}$ and $C_{ab}$ in eqs.(2.5) and (2.6) determine the
boundary conditions, we can impose fixed boundary conditions by
modifying the definition of $P_{abc}$ and $C_{ab}$. For example, if we
impose $s_i = 1$ for all $i$-site at the boundary, the boundary
weights become
\begin{eqnarray}
P_{abc}^{~} & = & {\rm exp}
\bigg\{ K^*( \delta_{1a} + \delta_{1b}
+ \delta_{1c}) \bigg\} \nonumber\\
C_{ab}^{~} & = & {\rm exp} \bigg\{ K^*( \delta_{1a}
+ \delta_{1b}) \bigg\}
\end{eqnarray}
according to the fixed boundary condition.

Now we can express the partition function $Z^L$ in terms of $P^N$ and
$C^N$. For even-size  ($L = 2N$) systems, the partition function is
written as
\begin{equation}
Z^{2N} = {\rm Tr~} \left( C^N \right)^4
= \sum_{{\mib a}{\mib b}{\mib c}{\mib d}}
C^N_{{\mib a}{\mib b}} \, C^N_{{\mib b}{\mib c}} \,
C^N_{{\mib c}{\mib d}} \, C^N_{{\mib d}{\mib a}} .
\end{equation}
For odd-size ($L = 2N+1$) cases, we generalize eq.(2.7) and express
$Z^{2N+1}$ as
\begin{equation}
Z^{2N+1}{\hspace{-1pt}} =  \sum_{kheb} W_{kheb}^{~} {\rm Tr}
\big( P^N_k C^N P^N_h C^N P^N_e C^N P^N_b C^N \big)
\end{equation}
where we have regarded $P^N$ as a matrix.

In addition to the partition function, we can obtain thermal averages
of the spin polarization or spin correlation functions in the same way.
For example, the $s$-spin at the center of the odd-size
($L = 2N + 1$) cluster takes the direction `1' with the probability
\begin{equation}
\langle \delta_{1s} \rangle = {
{\sum_{abcd} X_{abcd}^{~} {\rm Tr}
\big( P^N_a C^N P^N_b C^N P^N_c C^N P^N_d C^N \big) } \over
{\sum_{abcd} W_{abcd}^{~} {\rm Tr}
\big( P^N_a C^N P^N_b C^N P^N_c C^N P^N_d C^N \big) } } ,
\end{equation}
where $X$ is a new vertex weight
\begin{equation}
X_{abcd} = {\rm exp}
\bigg\{ K^*( \delta_{1a} + \delta_{1b}
+ \delta_{1c} + \delta_{1d} ) \bigg\}
\end{equation}
that counts the case $s = 1$.

The corner transfer matrix $C^N$ is $q^N$-dimensional, where the
dimension increases rapidly with $N$. The fact prevents us from exact
numerical calculation of $Z^L$ by way of eq.(2.12) and (2.13). For
example when $q = 2$, the upper limit of $N$ is about 13. The
restriction for $N$ is more sever for larger $q$ cases. This is the
main reason that we employ the renormalization group (RG) method.

\section{Minimum Free Energy Principle}

The background of DMRG is the minimum free energy (or the maximum
partition function) principle which is represented by the density
matrix. Generally speaking, a density matrix $\rho$ in statistical
mechanics is a matrix whose trace coincides with the partition
function of the system. The product of four CTMs
\begin{equation}
      {\rho}^{2N}_{~} =  \big(C^N\big)^4
\end{equation}
is a kind of density matrix, whose trace ${\rm Tr} \, \rho^{2N}$ is
the partition function $Z^{2N}$ in eq.(2.12). Figure 5 shows the
construction of $\rho^{2N}$ when $N = 3$; the element of the density
matrix $\rho^{2N}_{{\mib a}{\mib b}}$ represents the Boltzmann weight
of the system that has a cut (or gap), where ${\mib a}$ and ${\mib b}$
represent the spin configurations on each side of the cut.

\begin{figure}
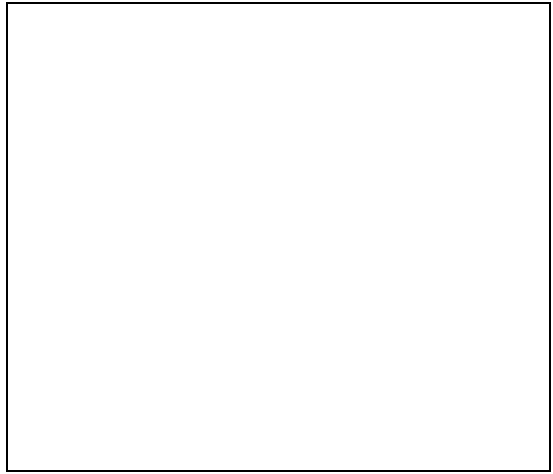

\figureheight{6cm}
\caption{Density matrix $\rho^{2N}$ for $N = 3$. The shaded region
represents the corner transfer matrix $C^3$.}
\label{fig:5}
\end{figure}

Equation (3.1) shows that both $C^N$ and $\rho^{2N}$ have the common
eigenvectors ${\mib R}_i$ $(1 \le i \le q^N)$ that satisfy the
eigenvalue equations
\begin{eqnarray}
C^N_{~} {\mib R}_i & = & {\mib R}_i \, \omega_i \nonumber\\
\rho^{2N}_{~} {\mib R}_i & = & {\mib R}_i \, \omega_i^4 \, ,
\end{eqnarray}
where $\omega_i$ and $\omega_i^4$ is the $i$-th eigenvalue of $C^N$
and $\rho^{2N}$, respectively. We write down the element of ${\mib
R}_i$ as $R_{{\mib a} j}^{~}$ in the following. According to the
symmetry of the $q^4$-vertex model, both $C^N$ and $\rho^{2N}$ are
symmetric. Therefore the square matrix $R \equiv ({\mib R}_1, {\mib
R}_2, \ldots)$ satisfies the orthogonal relation
\begin{eqnarray}
\big(R_{~}^{T} R\big)_{ij}^{~}
& = ( {\mib R}_i, {\mib R}_j )
& = \delta_{ij} \nonumber\\
\big(R R_{~}^{T}\big)_{{\mib a}{\mib b}}^{~}
& = \sum_i R_{{\mib a} i}^{~} R_{{\mib b} i}^{~}
& = \delta_{{\mib a}{\mib b}} .
\end{eqnarray}
We assume the decreasing order
$\omega_1^{~4} \ge \omega_2^{~4} \ge \cdots$ for the eigenvalues,
where all of the $\omega_i^{~4}$ are nonnegative; as far as our
$q^4$-vertex model is concerned, $\omega_i^{~2}$ is also real and
nonnegative. A trivial inequality
\begin{equation}
Z^{2N} \geq \sum_{\xi = 1}^m \omega_{\xi}^{~4}
\end{equation}
holds for arbitrary integer $m$; the l.h.s. coincides with the r.h.s.
when $m = q^N$, or when $\omega_i = 0$ for all $i > m$. (Hereafter we
use greek letters for indices that run from $1$ to $m$.) It has been
known that the dumping of $\omega_i^4$ with respect to $i$ is fairly
rapid.~\cite{Wh2,IRF,Bx3} Therefore the r.h.s. of eq.(3.4) is a good
approximation for $Z^{2N}$ for sufficiently large $m$. Typically $m$
is of the order of hundreds in realistic numerical calculations.

Let us rewrite eq.(3.4) into the matrix formula
\begin{equation}
Z^{2N} \geq {\rm Tr} \big( {\tilde R}{\tilde R}_{~}^T \rho^{2N} \big)
= \sum_{{\mib a}{\mib b}} \sum_{\xi = 1}^m
R_{{\mib b}\xi}^{~} R_{{\mib a}\xi}^{~} \rho^{2N}_{{\mib a}{\mib b}}
\end{equation}
where ${\tilde R}$ is the rectangular matrix
$({\mib R}_1, {\mib R}_2, \ldots, {\mib R}_m)$. The matrix product
${\tilde R}{\tilde R}_{~}^T$ is no more an identity matrix, but is a
kind of projection operator that satisfies
\begin{equation}
{\tilde R}{\tilde R}_{~}^T = \big({\tilde R}{\tilde R}_{~}^T\big)^2 ,
\end{equation}
where $m = {\rm rank} \, ({\tilde R}{\tilde R}_{~}^T)$ coincides with
${\rm Tr} \big({\tilde R}{\tilde R}_{~}^T\big)$. Let us consider a
$q^N$-dimensional matrix $\tilde I$ that satisfies ${\tilde I} =
{\tilde I}^2$ and ${\rm rank} \, ({\tilde I}) = m$. One finds ${\rm
Tr} \big( {\tilde R}{\tilde R}_{~}^T \rho^{2N} \big) \ge  {\rm Tr}
\big({\tilde I} \rho_{~}^{2N}\big)$, where the r.h.s. coincides with
the l.h.s. only when ${\tilde I} = {\tilde R}{\tilde R}_{~}^T$. If we
regard Eq.(3.5) as a variational relation for the partition function,
the projection operator ${\tilde R}{\tilde R}_{~}^T$ in eq.(3.5) gives
maximum variational partition function (or minimum free energy) under
the constraint ${\rm rank} \, ({\tilde R}{\tilde R}_{~}^T) = m$.

\begin{figure}
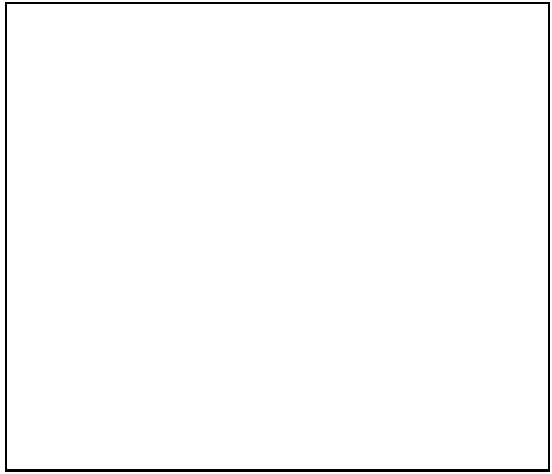

\figureheight{6cm}
\caption{Density matrix $\rho^{2N+1}$ for $N = 3$. Unlike $\rho^{2N}$,
the $\rho^{2N+1}$ is not symmetric.}
\label{fig:6}
\end{figure}

We can generalize the variational relation eq.(3.5) to finite size
systems with arbitrary shape. For example, let us consider an odd-size
($L = 2N + 1$) square cluster, which appears in eq.(2.13) and
eq.(2.15). We choose the example because it is a typical system whose
density matrix is  asymmetric.  The density matrix is defined as
\begin{equation}
\rho^{2N+1}_{{\mib a}{\mib b}} = \sum_{cdef} W_{cdef}
\big( P^N_c C^N P^N_d C^N
P^N_e C^N P^N_f C^N \big)_{{\mib a}{\mib b}} ,
\end{equation}
where arrangements of the spin indices are shown in Fig.6. We have
defined $\rho^{2N+1}$ so that the matrix dimension coincides with that
of $\rho^{2N}$ in eq.(3.1). Since $\rho^{2N+1}$ is asymmetric, the
formulation in eq.(3.2) and eq.(3.3) should be modified; we have to
consider the left eigenvalue problem
\begin{equation}
{\mib O}_i^T \rho^{2N+1}_{~} = \lambda_i^{~} {\mib O}_i^T
\end{equation}
independently from the right one
\begin{equation}
\rho^{2N+1}_{~} {\mib Q}_i^{~} = {\mib Q}_i^{~} \lambda_i^{~} \, ,
\end{equation}
where $\lambda_i$ is the eigenvalue in the decreasing order
$\lambda_1 \geq \lambda_2 \geq \cdots \geq 0$, and ${\mib O}_i$ and
${\mib Q}_i$ are the left and the right eigenvector of $\rho^{2N+1}$,
respectively. We express the vector element of ${\mib O}_i$ and
${\mib Q}_i$ as $O_{{\mib a} i}^{~}$ and $Q_{{\mib a} i}^{~}$,
respectively. This time the square matrices $Q \equiv ({\mib Q}_1,
{\mib Q}_2, \ldots)$ and  $O \equiv ({\mib O}_1, {\mib O}_2, \ldots)$
are not orthogonal by themselves, while they still satisfy the dual
orthogonal relation
\begin{equation}
\big(O^T Q\big)_{ij} = ( {\mib O}_i, {\mib Q}_j ) = \delta_{ij} \, .
\end{equation}
In other word, $O^T$ is the inverse of $Q$. The projection operator
for $\rho^{2N+1}$, which corresponds to ${\tilde R}{\tilde R}_{~}^T$ in
eq.(3.5), is then given by
\begin{equation}
\big({\tilde Q}{\tilde O}^T\big)_{{\mib a}{\mib b}}
= \sum_{\xi=1}^m Q_{{\mib a}\xi}^{~} O_{{\mib b} \xi}^{~} \, ,
\end{equation}
where ${\tilde Q} = ({\mib Q}_1, {\mib Q}_2, \ldots, {\mib Q}_m)$ and
${\tilde O} = ({\mib O}_1, {\mib O}_2, \ldots, {\mib O}_m)$ are
rectangular matrices. The dual orthogonal relation eq.(3.10) ensures
that the matrix $O^T Q$ satisfies $\big({\tilde Q}{\tilde O}^T\big)^2
= {\tilde Q}{\tilde O}^T$ and ${\rm Tr} \big({\tilde Q}{\tilde
O}^T\big) = m$. The inequality in partition function (eq.(3.5)) is
modified to
\begin{equation}
Z^{2N+1} \geq {\rm Tr} \big({\tilde Q}{\tilde O}^T \rho^{2N+1}\big)
\end{equation}
for the odd-size system.

We have obtained two projection operators, ${\tilde R}{\tilde R}^T$ in
eq.(3.5) and ${\tilde Q}{\tilde O}^T$ in eq.(3.11), where they have
the same matrix dimension ($= m$). Since ${\tilde R}{\tilde R}^T$
gives the maximum variational partition function when it is applied to
$\rho^{2N}$, we obtain the relation
\begin{equation}
{\rm Tr} \big({\tilde R}{\tilde R}^T \rho^{2N}\big) \geq
{\rm Tr} \big({\tilde Q}{\tilde O}^T \rho^{2N}\big) \, ,
\end{equation}
where the r.h.s. approaches to the l.h.s. with increasing $m$.
Conversely, we have
\begin{equation}
{\rm Tr} \big({\tilde Q}{\tilde O}^T \rho^{2N+1}\big) \geq
{\rm Tr} \big({\tilde R}{\tilde R}^T \rho^{2N+1}\big) .
\end{equation}
Let us compare the even-size system in Fig.5 with the odd-size one in
Fig.6. Both of them have the same cut (or gap) of length $N = 3$, but
their system sizes are different. The inequalities eq.(3.13) and
eq.(3.14) show that each system has its own optimal projection
operator, that is not the optimal one of other systems. The ratio
\begin{equation}
{\rm Tr} \big({\tilde R}{\tilde R}^T \rho^{2N+1}\big) / \,
{\rm Tr} \big({\tilde Q}{\tilde O}^T \rho^{2N+1}\big)
\end{equation}
depends on how the density matrix catches the boundary effect.
If the system is off-critical (or massive), the
boundary effect is expected to disappear in the thermodynamic limit $N
\rightarrow \infty$, and therefore the ratio eq.(3.15) approaches to
unity with increasing $N$. Even for a system at criticality, it is
numerically observed that the ratio approaches to unity with
increasing $N$.

\section{Renormalization group algorithm}

The projection operator ${\tilde R}{\tilde R}^T$ in eq.(3.5)
restricts the degree of freedom of $\rho^{2N}$ down to $m$. Therefore
the operation of the matrix ${\tilde R}$ on $\rho^{2N}$ can be
regarded as a RG transformation. In this sense the $m$-dimensional
diagonal matrix
\begin{equation}
{\tilde \rho}^{2N}_{~} = {\tilde R}^T \rho^{2N} {\tilde R}
= {\rm diag}(\omega_1^4, \omega_2^4, \ldots, \omega_m^4) .
\end{equation}
is the renormalized density matrix, that satisfies the variational
relation
\begin{equation}
Z^{2N} \geq {\rm Tr} \, {\tilde \rho}^{2N}_{~}
= {\rm Tr} \big({\tilde R}{\tilde R}^T \rho^{2N}\big) \, .
\end{equation}
Since $\rho^{2N}$ is written as a product of four CTMs, the RG
transformation in eq.(4.2) can be naturally extended to CTM:
\begin{equation}
{\tilde C}^N = {\tilde R}^T C^N {\tilde R}
= {\rm diag}(\omega_1, \omega_2, \ldots, \omega_m) \, .
\end{equation}

There are several ways to define the RG transformation for HRTM.
A way is to use the variational relation for $Z^{2N+1}$ in
eq.(3.12). Since $\rho^{2N+1}$ contains $P_a^N$ as shown in eq.(3.7),
the rectangular matrix ${\tilde O}$ in eq.(3.12) transforms $P_a^N$ as
${\tilde O}^T P_a^N {\tilde O}$. However, we don't follow this way,
because the renormalized HRTM thus defined is not
consistent with the renormalized CTM in eq.(4.2). In order to avoid
the problem, we define the RG transformation for HRTM using another
variational relation
\begin{equation}
Z^{2N+1} \geq {\rm Tr} \big({\tilde R}{\tilde R}^T
\rho^{2N+1}\big) \, ,
\end{equation}
which is derived from eq.(3.12) and eq.(3.14). This time ${\tilde R}$
touches $P_a^N$, which exists inside $\rho^{2N+1}$. (See Eq.(3.7))
>From eq.(4.4), we can define the renormalized HRTM as
\begin{equation}
{\tilde P}^N_a = {\tilde R}^T P^N_a {\tilde R} \, ,
\end{equation}
where the $(\xi\eta)$ element of ${\tilde P}^N_a$ is ${\tilde
P}^N_{{\xi}a{\eta}}$.

\begin{figure}
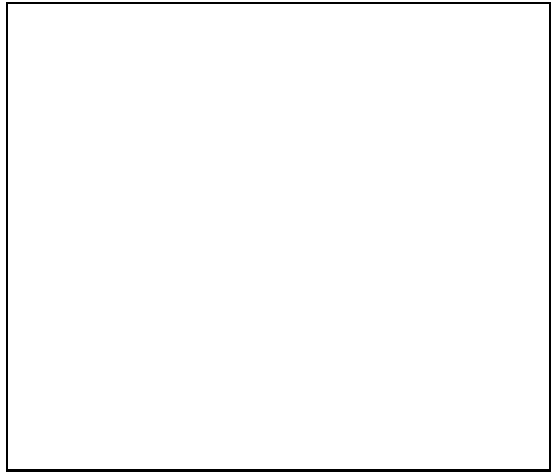

\figureheight{6cm}
\caption{Extension of the renormalized CTM. Note that
${\tilde C}^N$ is
a diagonal matrix.}
\label{fig:7}
\end{figure}

Now we explain the detail of our RG algorithm, that extends the
system size recursively using a mapping from ${\tilde C}^N$ and
${\tilde P}^N_a$ to  ${\tilde C}^{N+1}$ and ${\tilde P}^{N+1}_a$,
respectively. Remember that $C^N$ is defined recursively in eq.(2.8).
It is possible to generalize the recursion relation to the
renormalized CTM in eq.(4.3)
\begin{equation}
{\bar C}^{~N+1}_{(\alpha,a)(\beta,b)}
= \sum_{ef \delta} W_{efba}^{~} \, {\tilde P}^N_{\alpha e \delta}
{\tilde P}^N_{\beta f \delta} {\tilde C}^N_{\delta \delta} ,
\end{equation}
where ${\bar C}^{~N+1}_{(\alpha,a)(\beta,b)}$ is an extended CTM of
linear size $N+1$, and the pairs of indices $(\alpha,a)$ and
$(\beta,b)$ represent the row and the column indices of the
$qm$-dimensional matrix ${\bar C}^{N+1}$. (Fig.7) We have used the
fact that ${\tilde C}^N$ is diagonal: ${\tilde C}_{\alpha \beta}^N =
\delta_{\alpha \beta} \omega_{\alpha}$. The matrix ${\bar C}^{N+1}$ is
`partially diagonalized' in the sense that it contains $m$-state
block-spin indices $\alpha$ and $\beta$. We then perform the
renormalization group transformation on ${\bar C}^{N+1}$ and decrease
its matrix dimension ($= qm$) down to $m$. For this purpose we create
the density  matrix for the extended system
\begin{equation}
{\bar \rho}^{2(N+1)} = \big( {\bar C}^{N+1} \big)^4 \, ,
\end{equation}
and solve the eigenvalue problem
\begin{equation}
{\bar \rho}^{2(N+1)} \, {\bar {\mib R}}_i
= {\bar {\mib R}}_i \lambda_i
\end{equation}
in order to create a new RG transformation matrix ${\bar R} = ({\bar
{\mib R}}_1, {\bar {\mib R}}_2, \ldots, {\bar {\mib R}}_m)$, where the
eigenvalues $\lambda_i$ are in the decreasing order. As we have done
in eq.(4.3), we perform RG transformation for ${\bar C}^{N+1}$ as
\begin{equation}
{\tilde C}^{N+1} = {\bar R}^T {\bar C}^{N+1} {\bar R} \, ,
\end{equation}
to obtain the new renormalized CTM ${\tilde C}^{N+1}$, which is an
$m$-dimensional diagonal matrix. We also perform the RG
transformation for HRTM  at the same time. First we increase the
length of HRTM using the relation
\begin{equation}
{\bar P}^{N+1}_{(\alpha,a) b (\gamma,c)}
= \sum_{d=1}^q W_{adcb}^{~} {\tilde P}^N_{\alpha d \gamma}
\end{equation}
which is a generalization of eq.(2.8), and then renormalize
${\bar P}^{N+1}_b$ by applying ${\bar R}$ to ${\bar P}^{N+1}_b$:
\begin{equation}
{\tilde P}^{N+1}_a =  {\bar R}^T {\bar P}^{N+1}_a {\bar R} .
\end{equation}

In this way, we have obtained ${\tilde C}^{N+1}$ and ${\tilde
P}^{N+1}$ from ${\tilde C}^N$ and ${\tilde P}^N$ through
eqs.(4.6)-(4.11). This is a cycle in CTMRG. Since the matrix dimension
of ${\bar C}^N$ in eq.(4.6) is equal to (or less than) $qm$, we can
repeat the RG process up to arbitrary $N$. Compare to Baxter's
method~\cite{Bx1,Bx2,Bx3}, the process of the renormalization group
transformation (eq.(4.9)) is the same, but the way to extend CTM
(eq.(4.6)) is different; the chief difference is that the renormalized
HRTM appears explicitly in CTMRG, while it is absent in Baxter's
method.

The maximum matrix element in ${\tilde C}^N$ and ${\tilde P}^N$ grows
exponentially with respect to $N$, since free energy is extensive.
Therefore we should take an appropriate normalization for both
${\tilde C}^N$ and ${\tilde P}^N$ during the iteration. The simplest
way is just to divide these matrices by their largest element, and
store the normalization factor.

Every after the iteration, the variational (or the approximate)
partition function is obtained from the relations ${\tilde Z}^{2N} =
{\rm Tr} \big( {\tilde C}^N \big)^4$, or
\begin{equation}
{\tilde Z}^{2N+1} = \sum_{kheb} W_{kheb}^{~} {\rm Tr}
\big( {\tilde P}^N_k {\tilde C}^N {\tilde P}^N_h {\tilde C}^N
{\tilde P}^N_e {\tilde C}^N {\tilde P}^N_b {\tilde C}^N \big) \, .
\end{equation}
Thermodynamic functions can be calculated from the numerical
derivative of the variational free energy ${\tilde F} = - {\sl k}_B T
\, {\rm ln} \, {\tilde Z}^{2N}$ with respect to the temperature. Spin
polarization and correlation functions can be expressed as a product
of renormalized CTM and HRTM. For example, spin polarization at the
center of the odd-size system is calculated as
\begin{equation}
\langle \delta_{1s} \rangle = {
{\sum_{abcd} X_{abcd}^{~} {\rm Tr}
\big( {\tilde P}^N_a {\tilde C}^N {\tilde P}^N_b {\tilde C}^N
{\tilde P}^N_c {\tilde C}^N {\tilde P}^N_d {\tilde C}^N \big) } \over
{\sum_{abcd} W_{abcd}^{~} {\rm Tr}
\big( {\tilde P}^N_a {\tilde C}^N {\tilde P}^N_b {\tilde C}^N
{\tilde P}^N_c {\tilde C}^N {\tilde P}^N_d {\tilde C}^N \big) } },
\end{equation}
which is the renormalized expression for eq.(2.14). As shown in
eq.(4.12) and eq.(4.13), the renormalized system of size $L = 2N + 1$
has original $q$-state spin variables at the center. Therefore local
quantities at the center are most easily obtained.

The algorithm of CTMRG is closely related to the infinite system
algorithm in DMRG.~\cite{Wh1} Okunishi have investigated their common
thermodynamic limit.~\cite{Ok} The major difference between DMRG and
CTMRG is that the DMRG create density matrix using an eigenvector of
the row-to-row transfer matrix, while in CTMRG the density matrix is
expressed as a product of four CTMs. In other word, DMRG treats an
infinitely long $d = 2$ system of width $L$, while CTMRG treats a
square shaped finite size system of size $L$.

We have treated the symmetric $q^4$-vertex model as a reference
system. It is straightforward to apply the CTMRG to other systems,
such as the $J_1$-$J_2$ Ising model,~\cite{Kanamori,Ho1,Ho2} the IRF
model, and triangular systems. Speaking more generally, CTMRG is
applicable to periodic $d = 2$ classical lattice systems that have
short-range interactions and discrete spin (or site) variables. Since
the CTM of these models are not always symmetric, care must be taken
for the diagonalization of the density matrix; we don't explain the
detail here because it is rather lengthy.~\cite{com1}

\begin{figure}
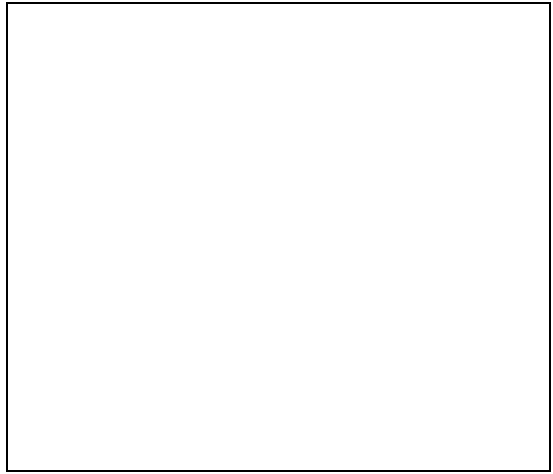

\figureheight{6cm}
\caption{A cut of length $3$ on the $d = 2$ lattice. The cut touches
the shaded 3 by 3 square block.}
\label{fig:8}
\end{figure}

In closing this section, we comment on the relation between the
conventional block RG~\cite{Kad} and CTMRG. Let us consider an
infinitely large square lattice that has a cut of length 3. (Fig.9) As
we have defined the density matrix for systems in Figs.5-6, we can
define $q^3$-dimensional density matrix that corresponds to the cut
(or gap) in Fig.9. The density matrix naturally leads a RG
transformation ${\tilde R}$ as we have mentioned in eq.(4.2), where
${\tilde R}$ transforms a set of 3 spins into an $m$-state block spin
variable. By applying the block spin transformation for every side of
the shaded square, we can map the 3 by 3 cluster into a $m^4$-vertex.
The CTMRG can be interpreted as a kind of block RG, where the block
size is equal to a quadrant of the whole system.

\section{Numerical Results}

The decorated Potts model, which we have treated as a $q^4$-vertex
model, has one-to-one correspondence to the Potts model on the simple
square lattice; see eqs.(2.1)-(2.3). Hereafter we refer to the latter
simply as `Potts model'. Using the correspondence, we perform
numerical calculations for $q^4$-vertex model and observe the scaling
behavior of the $q = 2, 3$ Potts model at criticality
\begin{equation}
K_c = {\rm ln} \, (\sqrt{q} + 1) .
\end{equation}
We examine the calculated data using finite size
scaling,~\cite{Fi,Ba} and compare the evaluated critical exponents
$\eta$ and $\nu$ with exact ones.

The order parameter of the $q$-state Potts model is defined
as~\cite{WuP}
\begin{equation}
M \equiv { {q\langle \delta_{1s} \rangle - 1} \over {q - 1} } .
\end{equation}
At the center of square cluster, it is expected
that the local order parameter $M$ obeys the scaling formula
\begin{equation}
M \sim L^{-(d-2+\eta)/2}
\end{equation}
at criticality, where $d$ is the spatial dimension $(=2)$, and $L$ is
the linear dimension of the system. We impose fixed boundary
conditions in eq.(2.11), because otherwise $\langle \delta_{1s}
\rangle$ is always zero. We evaluate $\langle \delta_{1s} \rangle$
using eq.(4.13).

\begin{figure}
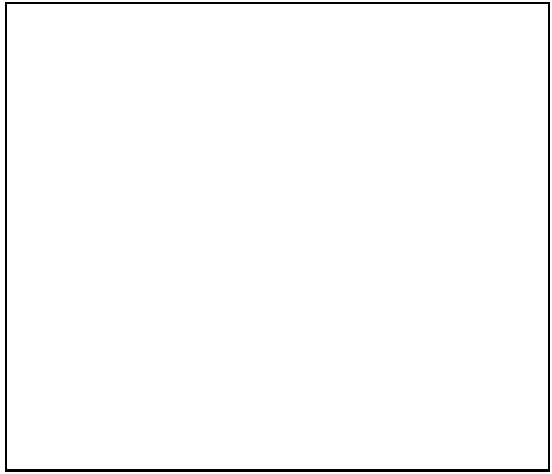

\figureheight{6cm}
\caption{Log-scale plot of the order parameter
$M$ in eq.(5.1) of the $q = 2$
Potts model.
}  \label{fig:9}
\end{figure}

Figure 9 shows the system size dependence of the order parameter $M$
in eq.(5.2) of the $q = 2$ Potts model, i.e., the Ising model. We plot
representative data when $m = 4$ and $m = 200$. The parameter $M$ is
almost proportional to $L^{-1/8}$, which is consistent with the
finite size scaling behavior in eq.(5.2) with $\eta = 1/4$. In fact,
the least-square fitting to the calculated data in the range $10 \leq
L \leq 1000$ gives $\eta = 0.2504$. We also perform the same scaling
analysis for the $q = 3$ Potts model. The obtained $\eta$ when $m =
200$ is summarized in Table I. The calculated exponents are in
accordance with the exact results.~\cite{WuP,CFT}

\begin{table}
\caption{Critical exponents $\eta$ of the $q = 2, 3$ Potts Models.
Least-square fittings are done in two regions,
$10 \leq L \leq 1000$ and
$100 \leq L \leq1000$.}
\label{table:1}
\begin{tabular}{@{\hspace{\tabcolsep}\extracolsep{\fill}}cccc} \hline
q & Exact & $L = 10 \sim 1000$  & $L = 100 \sim 1000$ \\ \hline
2 & $1/4 ~=$ 0.25 & 0.2504  & 0.2501 \\
3 & $4/15 ~=$ 0.2\.6 & 0.2652  & 0.2654 \\ \hline
\end{tabular}
\end{table}

We calculate another critical exponent $\nu$ using the finite size
scaling behavior of the local energy
\begin{equation}
E - E_c \sim L^{1/\nu -d}
\end{equation}
at criticality, where $E$ is the nearest
neighbor spin correlation function
\begin{equation}
E = \langle \delta(s_i, s_{i+1}) \rangle ,
\end{equation}
and $E_c$ is the local energy per site in the thermodynamic limit
$L \rightarrow \infty$. The value of $E_c$ is equal to~\cite{WuP,Bx3}
\begin{equation}
E_c = {1 \over 2} + {1 \over {2\sqrt{q}}} ,
\end{equation}
where the value comes from the duality relation.~\cite{WuP} Table II
shows the calculated exponents $\nu$ for $q = 2, 3$ Potts models when
$m = 200$. The calculated exponents agree with the exact
ones.~\cite{WuP,CFT}

One of the recent Monte Carlo (MC) simulation up to $N = 512$ by
Swendsen-Wang  algorithm~\cite{Sw} gives the exponent $\nu = 0.835(2)$
for the case $q = 3$.~\cite{Nw} Therefore, the numerical precision of
the CTMRG method is comparable to that of the recent MC simulation.
The superiority of the CTMRG is that the computation time is
proportional to $N$, while it is proportional to $N^2$ in MC
simulation. In addition, it should be noted that MC simulation
requires several independent runs in order to collect scaling data; in
CTMRG one can obtain the scaling data at once by single run.

\begin{table}
\caption{Critical exponents $\nu$ of the $q = 2, 3$ Potts Models.}
\label{table:2}
\begin{tabular}{@{\hspace{\tabcolsep}\extracolsep{\fill}}cccc} \hline
q & Exact & $L = 10 \sim 1000$  & $L = 100 \sim 1000$ \\ \hline
2 & 1              & 1.0017  & 1.0006 \\
3 & $5/6 =$ 0.8\.3 & 0.8323  & 0.8321 \\ \hline
\end{tabular}
\end{table}

We finally compare the numerical convergence in CTMRG to that in
Baxter's variational method.~\cite{Bx1,Bx2,Bx3} Figure 10 shows the
number of iterations that are necessary for obtaining site energy of
$q = 2$ potts model; we stop the iteration when the second largest
eigenvalue of the renormalized CTM converges down to the precision
$10^{-8}$ under the condition $m = 20$. The thick line shows the
iteration number by CTMRG, and the thin line shows that by Baxter's
variational method. Throughout the whole temperature region, CTMRG
exhibits better numerical convergence than Baxter's method. In
particular, at the temperature shown by the cross marks, the
calculation by Baxter's method is `trapped' at a quasi stable point,
and therefore the obtained results are not correct; such an
instability does not occur in CTMRG.

\begin{figure}
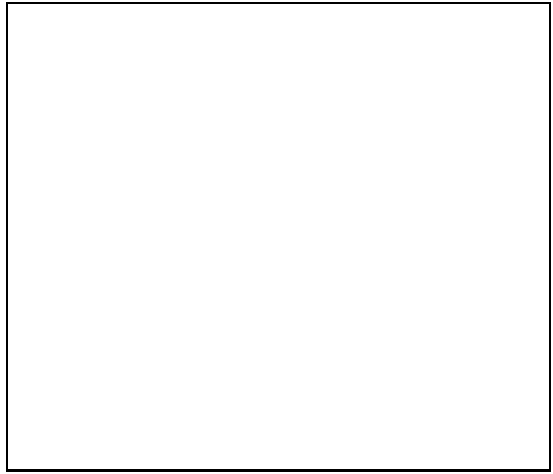

\figureheight{6cm}
\caption{Number of numerical iterations that are necessary for
obtaining site energy. The thick line shows the number by
CTMRG, and the thin line shows that by Baxter's variational method.
At the point shown by the marks, the free energy did not obtained
correctly by Baxter's method.}
\label{fig:10}
\end{figure}

\section{Conclusion}

We have explained the minimum free energy principle,
which is the common back ground in both DMRG and Baxter's CTM method.
>From the variational view point, we introduce the concept of DMRG into
Baxter's method, and obtain a new RG method for $d = 2$ classical
lattice models. Apart from the original DMRG, the numerical algorithm
of our method (CTMRG) is stable even at the critical point.

In order to check the efficiency of CTMRG, trial calculations are
performed for $q = 2, 3$ Potts models at criticality. Calculated data
are analyzed  using the finite size scaling method, and it is
confirmed that critical exponents of these models are correctly
evaluated by CTMRG.

The authors would like to express their sincere thanks to Y.~Akutsu
and M.~Kikuchi for valuable discussions. T.~N. thank to G.~Sierra,
M.~A.~Mart\`\i n-Delgado, and S.~R.~White for helpful discussions
about DMRG. K.~O. is supported by JSPS Research Fellowships for
Young   Scientists. The present work is partially supported by a
Grant-in-Aid from Ministry of Education, Science and Culture of Japan.
Most of the numerical calculations were done by NEC SX-3/14R in
computer center of Osaka University.


\begin{thebibliography}{99}
\bibitem{Hr} The half-row transfer matrix is sometimes called as
`vertex operator'.
\bibitem{com1} The matrix operations for such asymmetric CTMs are
explained by Baxter in his textbook$^{39)}$ p. 380.
\end{thebibliography}
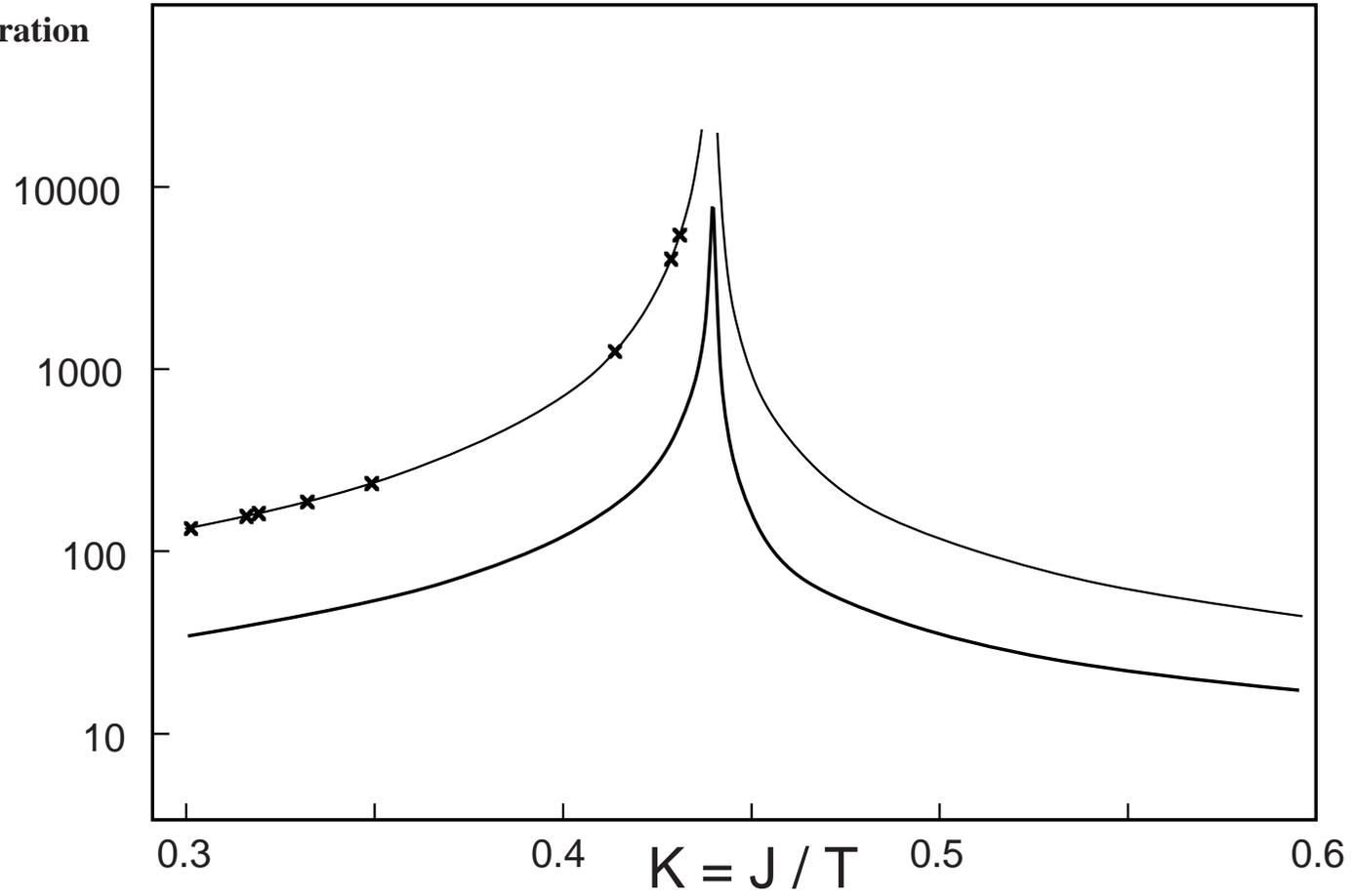
\end{document}